\documentstyle[12pt]{article}
\def \L{{\cal L}}
\renewcommand \a{\alpha}
\def \k{\kappa}
\def \Om{\Omega}
\def \l{\lambda}
\def \ov{\overline}
\def \pa{\partial}
\def \r{{\bf R}}
\def \A{A_0^{\k,q}}

\def \uA{\underline{A}}

\def \v{v^{\k,q}}

\def \e{\epsilon}
\def \F{{\cal E}}

\def \co{\frac{q_2\k_1}{q_1\k_2}}

\def \bb{\begin{equation}}
\def \ee{\end{equation}}
\def \bbs{\begin{eqnarray}}
\def \ees{\end{eqnarray}}
\def \bqrn{\begin{eqnarray*}}
\def \eqrn{\end{eqnarray*}}
\newcommand{\qed}{\vrule width0pt\hfill \raisebox{-.3ex}{
   \frame{\phantom{\vrule height6pt width4pt depth4pt}}} \hspace*{-
7pt}}
\newtheorem{Thm}{Theorem}  
\newtheorem{Def}{Definition}
\newtheorem{Lemma}{Lemma}
\newtheorem{Corollary}{Corollary}
\newtheorem{Pro}{Proposition}
\newenvironment{pf}{\medskip\noindent{\it Proof:}\enspace}{\hfill \qed \newline \medskip}

\begin{document}

\title{Vortex Condensates in the Relativistic Self-Dual Maxwell-Chern-Simons-Higgs System}
\author{Dongho Chae\\
Department of Mathematics \\
Seoul National University \\
Seoul, 151-742, Republic of Korea \\
e-mail address: {\em dhchae@math.snu.ac.kr}\\[1em]
Namkwon Kim\\
Basic Science Research Institute \& \\
Department of Mathematics \\
Seoul National University \\
Seoul, 151-742, Republic of Korea \\
e-mail address: {\em nkkim@math.snu.ac.kr}}
\date{ }
\maketitle
\abstract{The existence of vortex condensates in the self-dual 
Maxwell-Chern-Simons-Higgs System on a  flat torus is proved by the super-sub solution method
under the assumption that the total vortex number in  a given periodic domain is not too large. 
We also study the limiting behaviors of the solutions as the Chern-Simons coupling constant goes to some
limits. In the Abelian-Higgs limit we find that our solutions strongly converges to the corresponding vortex
condensates of the Abelian-Higgs system, while in the Chern-Simons limit the solutions strongly
converges to the corresponding vortex condensates of Chern-Simons system.  }

\section*{Introduction}
The Abelian-Higgs(or, Maxwell-Higgs) model was proposed by Ginzburg and Landau for phenomenological
study of the superconductivity. For the critical value of a parameter in the Lagrangian we obtain
a system of  self-duality equations, called the Bogomol'nyi equations, describing a static configuration
of vortices. This system of equations was studied rigorously by Jaffe-Taubes\cite{Taubes} for the case of whole domain of $\r ^2$. To explain the periodic array of vortices, called vortex condensates (pioneered by Abrikosov\cite{Ar}), Wang-Yang studied the same equations with
the periodic boundary condition in \cite{WY}. The vortices in this model has only magnetic charges. To have a theory
for vortices having both the electric and the magnetic charges Hong-Kim-Pac\cite{H} and Jackiw-Weinberg\cite{J} proposed
the Chern-Simons model(See also \cite{D}). The existence of topological multivortex solutions of the corresponding self-duality equations  in $\r ^2$ was proved
by \cite{W} using the variational argument similarly to \cite{J}, and later Spruck-Yang\cite{SY} constructed the solutions using a constructive
iteration scheme, and investigated  more detailed properties of solutions. For the Chern-Simons system with the periodic boundary
condition Caffarelli-Yang constructed a solution in \cite{CY} under the assumption that total vortex number in the given
 domain is not too big, and Tarantello\cite{T} refined the results of \cite{CY}, and proved, in particular,
multiplicity of solutions for some range of the Chern-Simons constant. 
In order to make a "unified theory" of the Abelian-Higgs and the Chern-Simons models C. Lee, K. Lee and
H. Min\cite{LLM1}(See also \cite{LLM2},\cite{LLW} and \cite{D}) suggested the self-dual Maxwell-Chern-Simons-Higgs theory using so called $N=2$ supersymmetry argument. 
For this theory the existence and various asymptotic properties of  topological multivortex solutions were
studied by the authors of this paper in \cite{CK}. In particular, the solutions constructed in \cite{CK} have the  properties that
in the limit of the Chern-Simons coupling constant $\kappa$ going to to zero with the electric charge $q$ kept
fixed(the Abelian-Higgs limit) the solutions converge strongly to the Abelian-Higgs vortices constructed
in \cite{Taubes}, while in the  limit $\kappa , q \rightarrow \infty$ with $2q^2/\kappa$ kept fixed the sequence of  parametrized solutions becomes ``weakly
consistent " to the Chern-Simons equations.  Due to this weak convergence of solutions in the Chern-Simons
limit the problem of  rigorous justification that the self-dual Maxwell-Chern-Simons-Higgs model is really a  unified theory
of  both of the Abelian-Higgs model and the Chern-Simons model  was not solved completely in \cite{CK}. \\
In this paper we study the vortex condensates of the self-dual Maxwell-Chern-Simons-Higgs model in a periodic
domain of $\r ^2$. We construct a solution using the super-sub solution method under the assumption that 
the total number of the vortices is not too big. We also study the Abelian-Higgs limit and the Chern-Simons
limit, and prove that our solutions converges {\em strongly} to the solutions of the Abelian-Higgs {\em and} the
Chern-Simons vortex condensates respectively. Moreover, since our arguments
leading to the strong convergence in Chern-Simons limit do not depend on our choice of periodic boundary condition,  and also works for the topological solutions in $\r ^2$ we resolve
the problem of obtaining convergence in this limit for the domain $\r ^2$ left unsolved in \cite{CK}. We thus complete the rigorous 
confirmation that
the self-dual Maxwell-Chern-Simons-Higgs model is, in some sense, an "interpolation" of the Abelian-Higgs and the Chern-Simons models both in the periodic domain and in  the whole domain of $\r ^2$.  \\
The organization of this paper is the following: In Section 1 we introduce the Lagrangian for the relativistic self-dual Maxwell-Chern-Simons-Higgs theory,  set up a system of semilinear partial differential equations resulting from the
corresponding Bogomol'nyi equations, and introduce the notion of admissible solutions following \cite{CK}.
In Section 2 we construct a subsolution under an assumption of relation among the Chern-Simons coupling
constant, electronic charge, and the total  vortex number. In Section 3 we prove existence of
an admissible solution using an iteration scheme. Finally in Section 4 we prove the strong convergence of our
admissible solutions to the Abelian-Higgs and the Chern-Simons solutions in the corresponding 
limits respectively.

\section{Preliminaries}
The Lagrangian for the self-dual Maxwell-Chern-Simons-Higgs theory  in \cite{LLM1} is 
\bbs   
\ov \L = (\ov D_{\mu} \ov \phi)(\ov D_{\mu}\ov \phi)^*  
       + \frac{1}{4}\ov F_{\mu\nu}\ov F^{\mu\nu}  
       + \frac{1}{4}\e^{\mu\nu\l}\ov A_{\mu}\ov F_{\nu\l}  
       - \frac{1}{2}(\ov \pa_\mu \ov N)^2 
                 \nonumber      \\
       - q^2 \ov N^2|\ov \phi|^2  
       - \frac{1}{2}(q |\ov \phi|^2 - \ov \k \ov N -q a^2)^2 
   \label{Lag}
\ees
Here we are considering the space-time domain  $\ov \Om \times \r \subset \r^2 \times \r$ 
with the metric tensor 
given by $g_{\mu \nu} =g^{\mu \nu} =diag(1, -1, -1)$, where $\ov \Om$ is a two dimensional flat torus.  
$\ov A = \ov A_{\mu}(\ov x) d\ov x^\mu$ is the gauge field,
  $\ov D_\mu = \frac{\pa}{\pa \ov x^\mu} - i \ov A_\mu$, $i=\sqrt{-1}$, is the gauge covariant derivative, 
$\ov F_{\mu\nu} = \ov \pa_\mu \ov A_\nu - \ov \pa_\nu \ov A_\mu$ is the curvature tensor, 
$ \e^{\mu\nu\l}$ is the skew symmetric tensor with $\e^{012}=1$, 
$\ov \phi =\ov \phi _1 +i\ov \phi _2 $ is a complex valued scalar field, called the Higgs field, $\ov N$
is the neutral scalar field, $\ov \k > 0$ is the Chern-Simons coupling constant, $q >0$ is  the  
charge of the electron, and  finally $\ov a$ is the symmetry breaking scale. In (1) the second and the third terms
are called the Maxwell term and the Chern-Simons term respectively. Given  $b>0$, let us consider the following
scale  transformation;
\bb  \label{trans} 
\ov A = b A  , \quad \ov x_\mu = \frac{ x_\mu}{b}  , \quad 
\ov \phi = b \phi , \quad \ov N = b N ,
\ee 
then the Lagrangian (\ref{Lag}) can be rewritten 
\bbs  
\L = \frac{\ov \L}{b^4}  
       = ( D_{\mu} \phi)( D_{\mu} \phi)^*  
          + \frac{1}{4} F_{\mu\nu} F^{\mu\nu}  
          + \frac{1}{4}\e^{\mu\nu\l} A_{\mu} F_{\nu\l} - \frac{1}{2}( \pa_\mu  N)^2  
      \nonumber \\ 
         -q^2  N^2| \phi|^2  
          -\frac{1}{2}(q |\phi|^2 -  \frac{\ov \k}{b}  N - q (\frac{a}{b})^2)^2 ,
      \label{Lag2}
\ees 
where $D_{\mu}$ and $F_{\mu\nu}$ are defined similarly to the above, using new variables $x_{\mu}$ and  new gauge field $A_{\mu}$, and  
$\ov \Om$ transforms into $\Om$ with the area given by $|\Om| = b^2|\ov \Om|$. 
From now on, we set $b =a$ and denote $\k = \frac{\ov \k}{a}$. 
The variational equations of $\ov \L$ on $\ov \Om \times \r$ are equivalent to 
those of $\L$ on $\Om \times \r$; 
\bbs \label{vari1} 
\frac{1}{2} \k \e^{\mu\nu\l}F_{\nu\l} + \pa^\nu F_{\mu\nu} = j^\mu  
            = i (\phi (D^\mu \phi)^* - \phi^* D^\mu \phi)  \\ 
     \label{vari2} 
\pa_\mu \pa^\mu N + 2 q^2 N |\phi|^2 + \k (q|\phi|^2 + \k N -q ) =0 \\ 
     \label{vari3} 
D_\mu(D^\mu \phi) + 2q^2N^2 \phi + q \phi (q|\phi|^2 + \k N -q ) =0 
\ees 
The Gauss  law (variational equation with respect to $A_0$) is given by  
\bb \label{Gauss} 
(- \Delta + 2q^2 |\phi|^2 )A_0 = -\k F_{12},
\ee 
while the static energy is
\bbs
\F = \int_{\Om} \biggl( |D_0 \phi|^2 + |D_j\phi|^2 + \frac{1}{2} F_{j0}^2  
         + \frac{1}{2}F_{12} + \frac{1}{2}(\pa_j N)^2   \nonumber \\
    \label{Energy}
         + q^2N^2|\phi|^2 + \frac{1}{2} (q|\phi|^2 + \k N -q )^2 \biggr) dx .
\ees

Since the system is invariant under the following gauge transformation 
\[ \phi \rightarrow e^{i\eta}\phi, \quad  
    A \rightarrow A + \nabla \eta, \quad N \rightarrow N	\] 
for any smooth real valued function $\eta$, 
$(\phi, A, N)$ satisfies the 't Hooft boundary conditions\cite{tH}; 
\bb \label{tH-con} 
\left. 
\begin{array}{c}    
e^{i\eta_k(x+\tau_k)} \phi(x+\tau_k) = e^{i\eta_k(x)}\phi(x) \\ 
(A + \nabla \eta_k)(x+\tau_k) = (A + \nabla \eta_k)(x)       \\ 
N(x+\tau_k) = N(x), \quad A_0(x+\tau_k) = A_0(x) 
\end{array} 
\right\} 
\ee 
Here, $\tau_k$, $k=1,2$ are the basis of the torus $\Om$. 
This equation leads to the following condition; 
\bbs      
\eta_1(1,1^-) - \eta_1(1,0^+) + \eta_1(0,0^+)- \eta_1(0,1^-) + \eta_2(0^+,1) 
                                      \nonumber \\
                - \eta_2(1^-,1) + \eta_2(1^-,0) - \eta_2(0^+,0) + 2\pi m = 0 
                \label{tH-sim}
\ees 
for an integer $m$.  Here, $(j,k)$, $j, k = 0^{\pm}, 1^{\pm}$  in the arguments stands for  
$j  \tau_1 + k \tau_2$. 
From  (\ref{Gauss}), (\ref{tH-con}), and (\ref{tH-sim}), we can  
obtain the quantized flux-charge relation as in the pure Chern-Simons  
system,
\[\Phi = \int_{\Om} F_{12} = 2\pi m, \quad  
   Q = \int j^0 = -2 \int q|\phi|^2A_0 = \frac{\k}{q} \Phi . \] 
The energy (\ref{Energy}) can be rewritten as follows by using the Bogomol'nyi type 
reduction;
\bbs 
\F &=& \int_{\Om} \left\{ |(D_1 \pm i D_2)\phi |^2 +|D_0\phi \mp iq\phi N|^2 
   +\frac{1}{2} (F_{j0} \pm \pa_j N )^2 \right. \nonumber \\ 
 && \ \ \ \ \left.+ \frac{1}{2}| F_{12} \pm (q|\phi|^2 +\kappa N -q )|^2  
           \right\}dx \pm q\int_{\r} F_{12}dx \nonumber \\ 
\label{BogoE} 
   && \mp \int_{\Om} \left( \nabla \cdot (N \nabla A_0)  
          + i [ \pa_1(\phi^*D_2\phi)-\pa_2(\phi^*D_1\phi)] \right) dx .
\ees 
Due to the 't Hooft boundary conditions (\ref{tH-con}) and (\ref{tH-sim}), the 
last term in (\ref{BogoE}) vanish after applying the integration by parts 
and $\int_{\r} F_{12}dx = 2\pi m$. Thus we have a lower bound of the energy 
\[ \F \geq 2\pi |mq| \]
and the following Bogomol'nyi equations which saturate the lower bound; 
\bbs 
A_0 = \mp N     \label{Bogo1}\\ 
(D_1 \pm iD_2)\phi = 0    \label{Bogo2}\\ 
F_{12} \pm (q|\phi|^2 + \k N - q ) = 0 \label{Bogo3} 
\ees 
From now on, we choose the upper sign of the above equations and  
assume $m \geq 0$. Indeed, for the negative $m$, we can obtain the solution  
$(\phi, A, N)$ by transforming simply the corresponding solution for $-m$.   
Following \cite{Taubes, WY}  we can apply $\overline{\pa}-$Poincar\'{e} lemma 
to (\ref{Bogo2}) 
to find that $\phi$ is analytic and has $m$ number of zeros counting 
multiplicities in $\Om$. 
Let $z_j$, $j=1,\cdots,k$ be the zeros of $\phi$ with multiplicities $m_j$ 
respectively. 
Due to the gauge invariance, we have a degree of freedom of  
the argument of $\phi$ up to smooth function and we can take 
\[ \phi = e^{\frac{u}{2} + i \theta} , \quad  
                    \theta = \sum_{j=1} ^k m_j Arg(z-z_j) \] 
by applying gauge transform if necessary.  Now, (\ref{Gauss}), 
(\ref{Bogo1}), (\ref{Bogo2}), and (\ref{Bogo3}) reduce to the followings; 
\bbs 
\Delta u &=& 2q^2 (e^u - 1) - 2q\k A_0 + 4\pi \sum_{j=1} ^k m_j\delta(z-z_j) \label{uraw} \\ 
\Delta A_0 &=& \kappa q (1 - e^u) + (\kappa^2 + 2q^2 e^u) A_0 \label{Araw} 
\ees 
Consider the equation 
\[ \Delta u_0 = 4 \pi \sum_{j=1}^k m_j \delta(z-z_j) - \frac{4\pi m}{|\Om|} \] 
The existence of $u_0$ is guaranteed and $u_0$ is smooth except $z = z_j$'s and 
behaves like $m_j \log (z-z_j)$ near $z=z_j$\cite{Au}.  We note that $u_0$  
is determined up to an additive constant.   
We set  $u= v + u_0$ in (\ref{uraw}) and (\ref{Araw}) to have the following equations with the Dirac delta
singularities removed; 
\bbs 
              \label{veq} 
\Delta v &=& 2q^2(e^{v + u_0} - 1 - \frac{\k}{q} A_0) + \frac{4\pi m}{|\Om|} \\ 
              \label{Aeq} 
\Delta A_0 &=& \k q (1 -e^{v + u_0}) + (\k^2 + 2q^2 e^{v+u_0})A_0 .
\ees 
If we formally set $\k =0$ in (\ref{veq}), then we obtain the equation for 
the Abelian-Higgs model in the periodic domain\cite{WY},
\bb 
              \label{ah} 
\Delta v = 2q^2(e^{v + u_0} - 1) + \frac{4\pi m}{|\Om|}.
\ee
On the other hand, the Chern-Simons equations with the coupling constant $1/l$ is
\bb
\label{cs} 
\Delta v =4l^2 e^{v+u_0} (e^{v + u_0} - 1) + \frac{4\pi m}{|\Om|}.
\ee
We denote the solutions of (\ref{ah}) and (\ref{cs}) by $v_a$ and $v^l _{cs}$ respectively; they are 
constructed in \cite{WY} and \cite{CY} respectively.
As in \cite{CK}, we introduce the following definition.
\begin{Def}
We call $(v, A_0) \in C^2 (\Om)$  an admissible solution of (\ref{veq}) and (\ref{Aeq})
if it is a solution of the equations satisfying $v + u_0 \leq 0$ and $A_0 \leq 0$.  
\end{Def} 
We note that the condition of admissibility is equivalent to 
$|\phi|^2 \leq 1$ and $N \geq 0$ in terms of $\phi$ and $N$, which is physically natural. 
In the pure Chern-Simons and the Abelian Higgs model this  
condition follows directly from the corresponding equations, using the maximum principle. This, however,   is not the case 
in our model. 
\begin{Pro} 
$(v, A_0)\in H^1(\Om)$ is admissible if and only if one of the followings hold; 
\begin{center}
\begin{enumerate} 
\item[(i)]  $\quad v  + u_0 \leq 0$ 
\item[(ii)]  $\quad A_0 \leq 0$ 
\item[(iii)]  $\quad v \leq v_a$ 
\item[(iv)]  $\quad \frac{q}{\k}(e^{v+u_0} -1) \leq A_0$.
\end{enumerate} 
\end{center}
\end{Pro} 
\begin{pf} 
From (\ref{Aeq}), if (i) holds, 
\[ \Delta A_0 \geq (\k^2 + 2q^2 e^{v+u_0})A_0    \] 
Thus, by applying the maximum principle, we have $A_0 \leq 0$, and  
$(v, A_0)$ is admissible. 
If (ii) holds, then from (\ref{veq}), we have
\[ \Delta (v+u_0 ) \geq 2q^2(e^{v+u_0}-1) = 2q^2 e^t (v+u_0 )    \] 
for some $t \in (v+u_0, 0)$ due to the mean value theorem.
Thus, (i) holds and $(v,A_0)$ is admissible. 
If (iii) holds, then $v + u_0 \leq 0$ since $v_a + u_0 \leq 0$\cite{WY}.   
Now, if (iv) holds, from (\ref{Aeq}), we obtain
\[ \Delta A_0 = \k^2 \{  \frac{q}{\k}(1 - e^{v+u_0}) + A_0 \} + 2q^2e^{v+u_0}A_0 
                               \geq 2q^2e^{v+u_0}A_0 .   \]
Thus again, by the maximum principle, we have (ii).
Conversely, if $(v, A_0)$ is admissible, then (i) and (ii) hold obviously.  
To show (iii), we note first  
\[ \Delta v_a = 2q^2 (e^{v_a + u_0} -1) + \frac{4\pi m}{|\Om|}.    \] 
By the mean value theorem we have 
\[ \Delta (v - v_a ) \geq 2q^2 (e^{v+u_0} - e^{v_a+u_0}) = 2q^2 e^t (v -v_a). \] 
Therefore, (iii) holds by the maximum principle.
Finally, by direct calculation using the fact $A_0 \leq 0$, we obtain the estimates
\bqrn
\Delta ( \frac{q}{\k}(1 - e^{v+u_0}) + A_0 ) 
          &\leq& - \frac{2q^3}{\k}e^{v + u_0}(e^{v+u_0} -1- \frac{\k}{q}A_0) \\
            && \quad  + q^2(1- e^{v+u_0}) +\frac{q}{\k}(\k^2 + 2q^2e^{v+u_0})A_0 \\
          &\leq& (\k q + 2q^2e^{v+u_0})(\frac{q}{\k}(1 - e^{v+u_0}) + A_0),
\eqrn
and (iv) holds by the mean value theorem again.
\end{pf} 
 
\section{Construction of  a subsolution} 
 
We say that $(w ,\uA)$ is a subsolution(supersolution) of  
(\ref{veq}) and (\ref{Aeq}) if
\bbs          \label{subdefv}
\Delta w &\geq(\leq)& 2q^2(e^{w + u_0} - 1 - \frac{\k}{q} \uA) + \frac{4\pi m}{|\Om|} \\
              \label{subdefA}
\Delta \uA &\geq(\leq)& \k q (1 -e^{w + u_0}) + (\k^2 + 2q^2 e^{w+u_0})\uA
\ees
We define an admissible subsolution(supersolution) similarly to an admissible solution.
The pair $( -u_0, 0)$ is obviously an admissible supersolution of (\ref{veq}) 
and (\ref{Aeq}).
To construct a subsolution of (\ref{veq}) and (\ref{Aeq}), we first define $f\in C^{\infty} (\Omega )$ by
\bbs     \label{deff}
f (z)= \left\{ 
\begin{array}{cc}
0 ,  & z\in \bigcup_{j=1} ^{k} B_{\delta}(z_j)  \nonumber \\
\a , &  z\in \Omega \setminus \bigcup_{j=1} ^{k} B_{2\delta}(z_j)  \nonumber \\
\frac{1}{2} \a (1+ \cos (\frac{\pi |z-z_j|}{\delta}) , & z\in B_{2\delta}(z_j) \setminus B_{\delta}(z_j) \
 ,
\end{array}
\right.
\ees
 where $B_{\delta}(z_j) =\{z\in \Omega \ | \ |z-z_j |< \delta \}$, and $z_j$'s are the zeros of $\phi$
with multiplicities $m_j$'s respectively.
Here, we have chosen 
$\delta = \frac{1}{2}\min_{i\neq j} \{ |z_i - z_j|, 1 \}$ so that each ball 
$B_{2\delta}(z_j)$ is disjoint with each other, and 
$ \a $ is the normalization constant to satisfy 
\[  \int_{\Om} f = |\Om| .  \]
Note that
\bb   \label{alpha}
1 \leq \a \leq \frac{|\Om|}{|\Om|- m \pi \delta^2} \leq 2, 
\ee
and 
\[ 
\begin{array}{cl}
    |\Delta f| \leq \left(\frac{\pi}{\delta} \right)^2 \a &
                      \mbox{if } \delta \leq |z-z_j| \leq  2\delta \ \mbox{for some $j$}    \\ 
    |\Delta f| = 0 & \mbox{otherwise} 
\end{array}
\]
As a candidate for a subsolution, we define $(w, \uA)$  as a pair of  smooth solutions of the system;
\bbs \label{subv} 
\Delta w &=& \frac{4 \pi m}{|\Om|}(1 - f) ,\quad \max_{z\in \Omega} (w(z) +u_0(z)) = u^* \\ 
     \label{subA} 
\uA &=& \frac{q}{\k}(e^{w+u_0} -1) + \frac{4\pi m}{2\k q |\Om|} f ,
\ees
where $u^* < 0$ is a constant to be determined in Lemma 1 below.  The existence of such $w$ is guaranteed  
by the fact that the integral  over $\Omega$ of the righthand side of (\ref{subv})
vanishes, and that $w$ is a solution determined up to an additive constant\cite{Au}. 
\begin{Lemma} 
The pair of functions, $(w, \uA)$, defined in (\ref{subv}) and (\ref{subA}), is a  
subsolution of (\ref{veq}) and (\ref{Aeq}) if  
\bb      \label{fcon1} 
\frac{4 \pi m}{|\Om|} \leq \frac{4q^4 e^{u^*}(1-e^{u^*})S}{\a( \k^2 + 4q^2e^{u^*} + (\frac{\pi}{\delta})^2) },
\ee
where $S =  \delta^{2m}e^{-C (1+\frac{|\log \delta|}{\delta}) m}$, 
$\delta$ is defined in the definition (\ref{deff}), and $C(\Om^0)$ is  
a constant depending only on the ratio of the side lengths of 
$\Om^0 = \frac{\Om}{|\Om|^{\frac{1}{2}}}=\{ x\in \frac{y}{|\Om|^{\frac{1}{2}}} \ | \ y\in \Omega\}$. 
\end{Lemma} 
\begin{pf} 
From the definitions of $w$ and $\uA$, (\ref{subdefv}) is satisfied  
obviously.  To show (\ref{subdefA}) holds, we start from 
\bqrn 
\Delta \uA &\geq& \frac{q}{\k}e^{w+u_0} \Delta(w+u_0)  
              + \frac{4\pi m}{2\k q |\Om|}\Delta f   \\ 
           &\geq& - \frac{4 \pi m q}{\k |\Om|} e^{w+u_0}f   
              + \frac{4\pi m}{2\k q |\Om|}\Delta f  \equiv LHS, 
\eqrn 
while
\bqrn 
RHS &\equiv& \k q (1-e^{w + u_0}) + (\k^2 + 2q^2 e^{w+u_0})\uA   \\ 
    &=& \frac{2q^3}{\k}e^{w+u_0}(e^{w+u_0}-1)  
           + \frac{4\pi m}{2\k q |\Om|}( \k^2 + 2q^2e^{w+u_0})f 
\eqrn 
Thus, the inequality $LHS \geq RHS$ holds if  
\[  
       \frac{4\pi m}{|\Om|}\left( -\Delta f 
           + ( \k^2 + 4q^2e^{w+u_0})f  \right) 
       \leq 4 q^4 e^{w+u_0}( 1 - e^{w+u_0} )               \] 
This, in turn, holds  for $z\in \bigcup_{j=1} ^{k} B_{\delta}(z_j) $ by the condition 
$\max_{\Omega} (w+u_0) = u^* < 0$. 
Thus, from our choice of $f$,  $(w,\uA)$ is a subsolution if  
\[ \frac{4 \pi m}{|\Om|} \leq \frac{4q^4 T}{\a (\k^2 +4q^2e^{u^*}  
      + (\frac{\pi}{\delta})^2 )} ,  \]
where we set 
\[ T = (1- \max_{ z\in \Omega \setminus \cup_{j=1} ^{k} B_{\delta}(z_j)  }  e^{w+u_0}) 
    \min_{ z\in \Omega \setminus \cup_{j=1} ^{k} B_{\delta}(z_j)  } e^{w+u_0}.   \]
We are going to estimate $T$ from now on.  Due to our choice of $w$, we have
\[  1- \max_{z\in \Omega \setminus \cup_{j=1} ^{k} B_{\delta}(z_j)  }  e^{w+u_0} \geq 1-e^{u^*}. \]
Now, define $u_1$ by 
\[
u_1 (z) = \left\{ 
\begin{array}{cc}
\frac{\a-f}{\a} m_j \log |z-z_j|^2  , & \quad \mbox{if } \  z\in B_{2\delta}(z_j)  \\
		0              , & \quad      \mbox{otherwise}
\end{array}
\right.
\]
and set $h = w +u_0 - u_1$. Note that $u_1 \leq 0$.  We obtain
\[ e^{w+u_0} = e^{u_1} e^h \geq \delta^{2m} e^h  \quad \mbox {on } \; 
z\in \Omega \setminus \cup_{j=1} ^{k} B_{\delta}(z_j) ,\] 
since $\delta \leq \frac{1}{2}$, and  
$m_j \leq m$.
We have  $\max_{\Om} h \geq \max_{\Om} (w + u_0) =  u^*$, since $u_1 \leq 0$.  Combining these facts, we have 
\[\min_{\Omega \setminus \cup_{j=1} ^{k} B_{\delta}(z_j)  } e^{w+u_0} \geq \delta^{2m} \exp (u^* 
                               - \| h \|_{osc(\Om)} ), \] 
where we denote  
\[ \| h \|_{osc(\Om)} = \sup_{x, y \in \Om} |h(x)-h(y)|    \] 
To estimate $\| h \|_{osc(\Om)}$, we first calculate 
\bb     \label{heq} 
 \Delta h = - \frac{4 \pi m}{|\Om|}f + K,  
\ee 
where we set   
\[ K = \left\{ 
     \begin{array}{cl}
        \frac{4 m_j \nabla f \cdot (z-z_j)}{\a |z-z_j|^2}  
              + \frac{2m_j}{\a} \log |z-z_j| \Delta f  
      & \mbox{ for }   \delta < |z-z_j| < 2 \delta \\
        0  & \mbox{ otherwise.}
              \end{array}
       \right.
\] 
By direct calculation we obtain 
\[ |\frac{\nabla f}{|z-z_j|}| \leq \frac{\a \pi}{2\delta^2}  , \quad 
      | \Delta f \log |z-z_j| | \leq \a (\frac{\pi}{\delta})^2|\log \delta| .  \] 
Thus, 
\bbs    
\| K \|_{L^2(\Om)} &\leq&  \sum_{j=1} ^k \frac{2m_j}{\a}  
            \left( 2 \left\| \frac{\nabla f}{|z-z_j|} \right\|_{L^2(\Om)} 
                + \| \Delta f \log |z-z_j| \|_{L^2(\Om)} \right)   \nonumber \\ 
\label{Kest} 
             &\leq& C m \frac{|\log \delta|}{\delta}, 
\ees 
where $C$ is an absolute constant. 
Now, using the coordinate transformation,  
$x \in \Om = |\Om|^{\frac{1}{2}} y $, $y \in \frac{\Om}{|\Om|^{\frac{1}{2}}}=\Om^0$, 
(\ref{heq}) becomes 
\[ \Delta h(y) = - 4 \pi m f(y) + |\Om| K(y)   \] 
on $\Om^0$ with $|\Om^0| = 1$. 
Thanks to Morrey's  imbedding inequality,  we obtain
\[ \|h\|_{C^\frac{1}{2}(\Om^0)} \leq C(\Om^0) \| D^2 h \|_{L^2(\Om^0)}  
                = C(\Om^0) \| \Delta h \|_{L^2(\Om^0)},             \] 
where $C(\Om^0)$ depends only on $\Om^0$. Therefore, 
\bqrn 
\| h \|_{osc(\Om)} &\leq& C(\Om^0) (diam(\Om^0))^{\frac{1}{2}} 
                                  \| \Delta h \|_{L^2(\Om^0)}  \\
                     &\leq& C(\Om^0) 4 \pi m \a + C m \frac{|\log \delta|}{\delta}   \\
                     &\leq& C(\Om^0)(1+ \frac{|\log \delta|}{\delta}) m
\eqrn 
by the estimate (\ref{Kest}) and the fact $\| |\Om| K(y) \|_{L^2(\Om^0)} =  
\| K(x) \|_{L^2(\Om)}$. Here, we denote various constants depending 
on the ratio of sidelengths of $\Om$ by $C(\Om^0)$. 
Therefore,  
\[ T \geq \delta^{2m}e^{u^*}(1-e^{u^*}) e^{-C(\Om^0)(1+\frac{|\log \delta|}{\delta}) m},    \] 
and thus  the proof of the lemma is completed by (\ref{fcon1}). 
\end{pf} 

\section{Existence of an admissible solution} 
 
Following \cite{CK}, we  define a sequence $(v^i, A^i_0)$, $i=0,1,2, \cdots$ inductively by
\bbs 
 && v^0 = -u_0, \quad A_0^0 = 0   \label{it1} \\ 
 (\Delta -d ) v^i &=& 2q^2 (e^{v^{i-1}+ u_0} - 1 - \frac{\k}{q}A_0^{i-1})  
                     + \frac{4 \pi m}{|\Om|} - d v^{i-1}     \label{it2}     \\ 
(\Delta -d ) A_0^i &=& \k q (1 - e^{v^i + u_0})  
                  + (\k^2 + 2q^2e^{v^i + u_0})A_0^i -d A_0^{i-1} 
                                     \label{it3} 
\ees 
for $d \geq 2q^2$.  
\begin{Lemma} 
Let $(w, \uA)$ be any admissible subsolution  pair of (\ref{veq}) - (\ref{Aeq}), 
and $(v^i, A_0^i)$ be defined as in (\ref{it1}), (\ref{it2}), 
and (\ref{it3}). Then $(v^i, A_0^i)$ is monotone decreasing with  
$i$ and satisfies 
\bb        \label{lowb} 
 v^i \geq w  , \quad A^i_0 \geq \uA  \ \ \ \forall i=0,1,2 \cdots  
\ee 
\end{Lemma} 
\begin{pf} 
We will prove only the inequality (\ref{lowb}) by the induction argument.  
The monotonicity  
can be proved by repetition of the proof of  Lemma 1 in \cite{CK}.  
From the admissibility condition, 
we obviously have $v^0 \geq w$ and $A^0_0 \geq \uA$.  Assuming (\ref{lowb}) holds 
for $i$, we have, by subtracting (\ref{subdefv}) from (\ref{it2}),  
\bqrn  
(\Delta -d) (v^{i+1}-w)  
               &\leq& 2q^2 (e^{v^{i} + u_0} -e^{w + u_0})  
                            - 2 \k q (A_0^{i} - \uA) - d (v^i - w)      \\ 
               &\leq& 2q^2 (e^{v^{i} + u_0} -e^{w + u_0}) - d (v^i - w) \\ 
               &\leq& (2q^2 e^\l -d )(v^i - w) 
\eqrn 
for some $\l \in (w+u_0, v^i + u_0)$, using  the mean value theorem.
From the monotonicity of $v^i$,  
$v^i + u_0 \leq 1$, which implies  
\[       (\Delta -d) (v^{i+1}-w) \leq 0.   \]  
Thus, applying the maximum principle, we have $v^{i+1} \geq w$. 
Again by subtracting the inequality (\ref{subdefA}) from (\ref{it3}), 
we have 
\bqrn 
(\Delta -d) (A_0^{i+1}-\uA)  
             &\leq& -\k q (e^{v^{i+1} + u_0} - e^{w + u_0})    \\
             &&  + (\k^2 + 2q^2e^{v^{i+1} + u_0})(A_0^{i+1}-\uA)   \\ 
             &&   + 2q^2(e^{v^{i+1} + u_0}-e^{w + u_0})\uA -d(A_0^i-\uA) \\ 
             &\leq& (\k^2 + 2q^2e^{v^{i+1} + u_0})(A_0^{i+1}-\uA),
\eqrn
 using the result, $v^{i+1} \geq w$. 
Thus, applying the maximum principle once more, we obtain 
$A_0^{i+1} \geq \uA$. This competes  the proof. 
\end{pf} 
 
\begin{Thm} 
Given $z_j \in \Om$ and nonnegative integers $m_j$, $j=1,\cdots, k$ with
$\sum_j m_j = m \in Z^+$, there exists a constant $C$ depending only on the  
ratio of the sidelengths of $\Om$ such that  
there exists a smooth admissible  energy minimizer 
with zeros  at $z =z_j$'s of multiplicity 
$m_j$ if $m$ satisfies the condition (\ref{fcon1}).
\end{Thm}
\begin{pf} 
The existence of such minimizer is guaranteed if 
we have an admissible solution of (\ref{veq}) and 
(\ref{Aeq}).  By  Lemma 3, it suffices to prove existence of  an admissible subsolution 
pair $(w, \uA)$.  
Now, we consider $(w, \uA)$ defined in (\ref{subv}) and (\ref{subA}) 
with $u^*$ chosen later.  By  Lemma 1, $(w, \uA)$ is a subsolution 
if (\ref{fcon1}) is satisfied.  
$w \leq -u_0 $ by the definition (\ref{subv}).  Thus, the condition $\uA \leq 0$ 
is enough to guarantee that  $(w, \uA)$ is admissible which reads 
\[ 
\frac{q}{\k}(e^{w+u_0} - 1) + \frac{4\pi m}{2\k q |\Om|} f \leq 0  \]
Thus, it is enough 
\bb
 \frac{4\pi m}{|\Om|} \leq 2q^2 \min_{ \Om} \frac{1 - e^{w+u_0}}{\a}  
                                      =  2q^2  \frac{1-e^{u^*}}{\a} 
                                   \label{fcon2} 
\ee 
The condition (\ref{fcon2}), in turn,  follows immediately from (\ref{fcon1}).
\end{pf} 
\ \\
\noindent{\bf Remark:} 
The solution we have constructed  under the assumptions on $\frac{4 \pi m }{|\Om|}$ in Lemma 1 is
maximal among admissible solutions.  
Thus it describes the most superconducting state.  \\
\ \\
We now establish some ordering properties among the admissible
solutions.
\begin{Thm}
Given $z_j$ and $m_j$, $j=1, \cdots, k$  as in  Theorem 1, 
let  $(v^{\k_1, q_1}, A_0^{\k_1, q_1})$ be an admissible solution for $\k=\k_1$ and $q=q_1$, then
\begin{enumerate}
\item[(i)] $(v^{\k_1,q_1}, \frac{\k_1}{\k_2}A_0^{\k_1,q_1})$ is a subsolution for $\k=\k_2 < k_1$ and
                 $q=q_1$.
\item[(ii)] $(v^{\k_1,q_1}, \frac{q_1}{q_2}A_0^{\k_1,q_1})$  is a subsolution for $q=q_2 > q_1$ and
            $\frac{q^2_1}{\k_1} = \frac{q^2_2}{\k_2}$.
\item[(iii)] $(v^{\k_1,q_1}, \frac{q_2\k_1}{q_1\k_2}A_0^{\k_1,q_1})$  is a subsolution for
                $q=q_2 \geq q_1$ and $\k=\k_2 \leq k_1$.
\end{enumerate}
\end{Thm}
\begin{pf}
Note that (i) and (ii) are  special cases of (iii).  Thus, it suffices to  prove (iii) only. 
From (\ref{veq}) we have
\bqrn
\Delta v^{\k_1, q_1} &\hspace{-1em}-&\hspace{-1em} 2q_2^2(e^{v^{\k_1, q_1} + u_0} - 1 
                            - \frac{\k_2}{q_2} A_0^{\k_1, q_1}) - \frac{4\pi m}{|\Om|}   \\
                 &=& 2(q_1^2-q_2^2)(e^{v^{\k_1, q_1} + u_0} - 1- \frac{\k_1}{q_1}A_0^{\k_1, q_1}) \\
							&&			- 2(\frac{\k_1}{q_1}q_2^2 - \k_2 q_2)A_0^{\k_1, q_1} \\
                 &\geq& 2q_2^2A(\frac{\k_2}{q_2} - \frac{\k_1}{q_1}) \geq 0,
\eqrn
since $A_0 \leq 0$ and $e^{v^{\k_1, q_1} + u_0} - 1- \frac{\k_1}{q_1}A_0^{\k_1, q_1}$.
Now, from (\ref{Aeq}) we  obtain for $q=q_2 \geq q_1$ and $\k=\k_2 \leq k_1$,
\bqrn
\Delta \co A_0^{\k_1,q_1} &\hspace{-1em}-&\hspace{-1em}  \k_2 q_2 (1 -e^{v^{\k_1,q_1} + u_0}) 
                            + (\k_2^2 + 2q_2^2 e^{v^{\k_1,q_1}+u_0}) \co A_0^{\k_1,q_1}    \\
             &=& (\frac{\k_1^2 q_2}{\k_2} - \k_2 q_2) (1 -e^{v^{\k_1,q_1} + u_0})   \\
                         &&   + (\k_1^2 - \k_2^2 + 2(q_1^2 - q_2^2) e^{v^{\k_1,q_1}+u_0})\co A_0^{\k_1,q_1} \\
             &\geq&  (\frac{\k_1^2 q_2}{\k_2} - \k_2 q_2)(1 -e^{v^{\k_1,q_1} + u_0} 
                                - \frac{\k_1}{q_1}A_0^{\k_1,q_1}) \geq 0
\eqrn
This completes the proof of Theorem 2.
\end{pf}
\begin{Corollary}
Denoting the maximal admissible solution for $\k, q$ by $(v^{\k,q}, A_0^{\k,q})$, we have
\begin{enumerate}
\item[(i)] $v^{\k,q} \leq v^{\k',q}$, $\k A_0^{\k,q} \leq \k' A_0^{\k',q}$ for $\k' < \k$.
\item[(ii)] $v^{\k,q} \leq v^{\k',q'}$, $q A_0^{\k,q} \leq q' A_0^{\k',q'}$ for $q' > q$, 
                    $\frac{q^2}{\k} = \frac{q'^2}{\k'}$.
\item[(iii)] $v^{\k,q} \leq v^{\k',q'}$, $\k q' A_0^{\k, q} \leq \k' q A_0^{\k',q'}$ 
               for $\k' \leq \k$   and $q' \geq q$.
\end{enumerate}
\end{Corollary}
{\bf Remark:}
We note that Theorem 2 and Corollary 1 hold also  for $\Om = \r^2$,  since the argument of proof does not depend
on our choice periodic boundary condition, and the maximum principle used above also holds for topological solution in
 $\Om = \r^2$.  
This observation is important in the remark at the end of the next section.\\
\ \\
As an application of Theorem 2 we have the following:
\begin{Thm}
Given $z_j$ and $m_j$, $j=1, \cdots, k$  as in  Theorem 1, there exist 
 critical constants, $0 < \k_c < 1$ and $\infty > q_c > 1$ such that the following holds:
An admissible solution of (\ref{veq}) and (\ref{Aeq}) exists
if
\[  \frac{4 \pi m}{|\Om|} < \min \{ \k_c^2\frac{q^4}{\k^2}, \;\;  2\frac{q^2}{q_c^2} \},      \]
and does not exist
if
\bb         \label{nncon2}
 \frac{4 \pi m}{|\Om|} > \min \{ \k_c^2\frac{q^4}{\k^2},  
                             2\frac{q^2}{q_c^2} \}.    
\ee
\end{Thm}
\begin{pf}
Assume that  there exists an admissible solution. Then, integrating (\ref{veq}), we have
\[
0 = \int_{\Om} \Delta v = -2q^2 \int_{\Om} ( 1 - e^{v+u_0} + \frac{q}{\k}A_0)
    + 4\pi m    \]
Thus, using the fact $A_0 \leq 0$, and $A_0$ is not identically zero, we have
\bb     \label{ncon1}
\frac{4\pi m}{|\Om|} < 2 q^2.
\ee
Now integrating $\Delta(v + 2\frac{q}{\k}A_0)$, we have
\[    0 = \frac{4q^3}{\k}\int e^{v+u_0}A_0 + 4 \pi m.   \]
Then, using (iv) of the Proposition 1,
\[
4 \pi m \leq \frac{4q^4}{\k^2}\int e^{v + u_0}(1-e^{v+u_0}) < \frac{q^4}{\k^2}|\Om|
\]
by the inequality $t-t^2 \leq \frac{1}{4}$ for $0 \leq t \leq 1$.
Thus, we have
\bb      \label{ncon2}
\frac{4 \pi m}{|\Om|} < \frac{q^4}{\k^2}.
\ee
Therefore, there are constants $k_c < \infty$ and $q_c > 0$ such that the condition (\ref{nncon2})
holds.
Now, by the condition (\ref{fcon1}) and Theorem 1, if $\frac{4 \pi m}{|\Om|}$ is sufficiently small compared to $\min \{ \frac{q^4}{\k^2}, 2q^2 \}$, then there is an admissible solution.  Therefore it is enough to prove that if for 
a certain $\k_1$ and $q_1$ there exists
an admissible solution,  then there exists an admissible solution for all $\k < \k_1$ and
$q > q_1$.  By  Lemma 2, it suffices to have an  admissible subsolution 
for $\k < \k_1$ and $q > q_1$.  Existence of such a subsolution is established in Theorem 2. This completes the proof  of 
Theorem 3.
\end{pf}

\section{The Abelian-Higgs and the Chern-Simons limits}

In this section we prove the strong convergence of admissible solutions both in the Abelian-Higgs limit and in the Chern-Simons limit. On the one hand, this proves rigorously that our model corresponds to an interpolation of those two models.
On the other hand,  combining with our existence theorem in the previous section, establishes existence of solutions
of the those model equations in different methods from \cite{WY} and \cite{CY}. We recall that $v_a$ and $v_{cs} ^l$ below
are the Abelian-Higgs solution and the Chern-Simons solution defined by the equations (\ref{ah}) and (\ref{cs}) respectively.
\begin{Thm}
Let $(\v, \A)$ be an admissible solution with zeros $z_j$ $m_j$, $j=1, \cdots, k$. Then,  we have
\begin{description}
\item ( The Abelian -Higgs limit); \\
$\v \rightarrow v_a$, and $\A \rightarrow 0$  both in $C^{\infty} (\Om)$ as $\k \rightarrow 0$ with $q$ kept fixed.
\item {The Chern-Simons limit};  \\
$\v \rightarrow v_{cs}^l$, and $\frac{\k}{q}\A \rightarrow e^{v_{cs}^l +u_0} -1$ both in $H^1 (\Om)$
 as $\k \rightarrow \infty$ with $\frac{q^2}{\k} = l$ kept fixed.
\end{description}
\end{Thm}
\begin{pf}
We first consider the Abelian-Higgs limit.
By (i) of Corollary 1 and  Proposition 1 we have
\[   |\A| \leq \k |A_0^{q,\k_0}| \leq \frac{\k q}{\k_0}   \]
Thus $\A \rightarrow 0$ in $L^{\infty}(\Om)$.
Subtracting the equation (\ref{ah}) for $v_a$ from the equation (\ref{veq}) for $\v$, we obtain
\bb     \label{vvh}     
\Delta (\v - v_a) = 2q^2(e^{\v+u_0} -e^{v_a + u_0}) - 2\k q \A .
\ee
Thus, $\| \Delta (\v - v_a) \|_{L^p(\Om)} \leq 4q^2 |\Om|^{\frac{1}{p}}$,  by 
Proposition 1.  Using the Cald\'{e}ron-Zygmund inequality and the compactness of the 
embedding $W^{2, p} (\Omega ) \hookrightarrow  C^1 (\Omega)$, $ p>2$, there exists a subsequence  $\{  \v \}$ such that
$\v \rightarrow v_a$ in $C^1 (\Om)$.  On the other hand , thanks to the monotonicity  (i) of Corollary 1, and the
admissibility, $v^{\k_0 ,q} \leq \v \leq v_a$, 
the original sequence  $\{  \v \}$  actually converges to $v_a$ in $C^1(\Omega)$. 
$C^{\infty} (\Omega ) $-convergence results from (\ref{vvh}), applying the bootstrapping argument combined with the 
standard elliptic regularity. \\
Now, we consider the Chern-Simons limit.
By (ii) of Corollary 1, we have
\[ 0 \geq \A  \geq \frac{q_0}{q}A_0^{\k_0, q_0},   \]
thus $\A \rightarrow  0$ in $L^{\infty}(\Om)$.
Since $\v$ is monotone increasing as $\k \rightarrow \infty$ and $\v \leq -u_0$, 
there exists a pointwise limit, $v_{cs}^l \in L^2(\Om)$ and 
\[ \lim_{\k, q \rightarrow \infty} \int |\v -v_{cs}^l |^2 = 0  \]
by the Lebesgue dominated convergence theorem.  
Applying the Lebesgue dominated convergence theorem again, we have
$ e^{\v+u_0}-1 \rightarrow e^{v_{cs}^l +u_0}-1$ in $L^p(\Om)$ for any $ 1 \leq p < \infty$.
Now, integrating the (\ref{veq}), we have
\[ 2q^2 \int_{\Om} (e^{\v+u_0} -1 - \frac{\k}{q}\A) + 4\pi m = 0    \]
Thus, by the admissibility, $\|e^{\v+u_0} -1 - \frac{\k}{q}\A\|_{L^1(\Om)} 
= \frac{4 \pi m}{2q^2} \rightarrow 0$ as $q \rightarrow \infty$.  Therefore,
$\frac{\k}{q}\A = \frac{q}{l}\A \rightarrow  e^{v_{cs}^l+u_0} -1$ in $L^p(\Om)$ 
for any $ 1 \leq p < \infty$.
Subtracting (\ref{Aeq}) after multiplication by $\frac{q}{\k}$ from (\ref{veq}), we have
\bb     \label{appcs}
\Delta ( \v + \frac{2q}{\k} \A ) = \frac{4q^3}{\k}e^{\v+u_0} \A + \frac{4 \pi m}{|\Om|}.  
\ee
Multiplying (\ref{appcs}) by a periodic test function $\phi \in C^2(\Om)$,  and taking the limit of the equation, we have
\[
LHS = \lim_{\k, q \rightarrow \infty} \int (\v + \frac{2q}{\k} \A) \Delta \phi = \int v_{cs}^l \Delta \phi \]
\bqrn
\lim_{\k, q \rightarrow \infty} \int \frac{4q^3}{\k}e^{\v+u_0} \A \phi 
          &=& 4 l^2 \lim_{\k, q \rightarrow \infty} \left( \int  e^{\v +u_0}(1- e^{\v+u_0}) \phi \right. \\
                &&   +   \left. \int e^{\v +u_0}(1- e^{\v+u_0} + \frac{\k}{q} \A) \phi \right) \\
          &=& 4 l^2 \int  e^{v_{cs}^l +u_0}(1- e^{v_{cs}^l+u_0}) \phi .
\eqrn
Thus $v_{cs}^l$ is a weak solution of the Chern-Simons equation.  Using the 
standard elliptic regularity theory,
we obtain $v_{cs}^l \in C^{\infty} (\Om)$.  Thus,  $v_{cs}^l$ is indeed a classical solution
of the Chern-Simons equation.   To show the convergence in $H^1(\Om)$,  we first subtract
the corresponding Chern-Simons equation from (\ref{appcs}) and integrating it
after multiplying by $\v -v_{cs}^l$ to get
\bqrn
\int |\nabla(\v-v_{cs}^l)|^2 &=& -\int  \frac{4q^3}{\k}e^{\v+u_0} \A(\v-v_{cs}^l)  \\
                    && + \int \Delta (\v-v_{cs}^l) \frac{q}{\k}\A \\
                      &\leq& C \int|\v-v_{cs}^l|  + C \frac{q}{\k} \rightarrow 0,
\eqrn
where we used 
\[  | \frac{4q^3}{\k}e^{\v+u_0} \A | \leq 4 l^2 | \frac{\k}{q} \A | \leq 4 l^2,  \]
and
\[ \int |\Delta (\v-v_{cs}^l)| \leq \int (|\Delta \v | + | \Delta v_{cs}^l| )\leq 16 \pi m. \]
This completes the proof of  Theorem 4.
\end{pf}
\ \\
\noindent{\bf Remark: } 
As noted in the remark after Corollary 1, since the Theorem 2  and Corollary 
also hold for $\Om =\r^2$ , and the above proof, using 
the monotonicity properties of Corollary 1 crucially, also work for  $\Om =\r^2$  with trivial modifications, we can deduce that the 
sequence of  admissible solutions of our system  converges in $H^1(\r^2)$  to  the Chern-Simons solutions in the Chern-Simons
limit. 
\newpage
\begin{center}
{\bf Acknowledgements}
\end{center}
The authors would like to thank Professor Choonkyu Lee for suggestion of the problem and encouragements.
The first author has been  supported partially by
GARC-KOSEF, BSRI-MOE, KOSEF(K95070),  and  SNU Research Fund, while the second author has been supported partially by KOSEF.

 \end{document}